\documentclass[]{spie}  

\usepackage{amsmath,amsfonts,amssymb}
\usepackage{graphicx}
\usepackage[colorlinks=true, allcolors=blue]{hyperref}

\newcommand{\hal}{\mbox{H$\alpha$}}

\title{The Prototype Telescope and Spectrograph System for the AMASE Project}

\author[a]{Renbin Yan}
\author[b,c,d]{Matthew A. Bershady}
\author[b]{Michael P. Smith}
\author[e]{Nicholas MacDonald}
\author[f,g]{Dmitry Bizyaev}
\author[e]{Kevin Bundy}
\author[c]{Sabyasachi Chattopadhyay}
\author[h]{James E. Gunn}
\author[e]{Kyle B. Westfall}
\author[b]{Marsha J. Wolf}

\affil[a]{Department of Physics and Astronomy, University of Kentucky, 505 Rose St., Lexington, KY, 40506 USA}
\affil[b]{Department of Astronomy, University of Wisconsin - Madison, 475 N. Charter Street, Madison, WI 53706-1582, USA.}
\affil[c]{South African Astronomical Observatory, PO Box 9, Observatory 7935, Cape Town, South Africa.}
\affil[d]{Department of Astronomy, University of Cape Town, Private Bag X3, Rondebosch 7701, South Africa.}
\affil[e]{UC Observatories, UCO/Lick, UC Santa Cruz, 1156 High St., Santa Cruz, CA 95064, USA}
\affil[f]{Apache Point Observatory and New Mexico State
University, P.O. Box 59, Sunspot, NM, 88349-0059, USA}
\affil[g]{Sternberg Astronomical Institute, Moscow State University, Moscow, Russia}
\affil[h]{Department of Astrophysical Sciences, Princeton University, 4 Ivy Lane, Princeton, NJ, 08544}

\authorinfo{Further author information: (Send correspondence to R.Y.)\\R.Y: E-mail: rya225@g.uky.edu, Telephone: 1 347 459 8709}

\begin{document} 
\maketitle

\begin{abstract}
We present the design of the prototype telescope and spectrograph system for the Affordable Multiple Aperture Spectroscopy Explorer (AMASE) project. AMASE is a planned project that will pair 100 identical multi-fiber spectrographs with a large array of telephoto lenses to achieve a large area integral field spectroscopy survey of the sky at the spatial resolution of half an arcminute and a spectral resolution of R=15,000, covering important emission lines in the optical for studying the ionized gas in the Milky Way and beyond. The project will be enabled by a significant reduction in the cost of each spectrograph unit, which is achieved by reducing the beam width and the use of small-pixel CMOS detectors, 50${\rm \mu m}$-core optical fibers, and commercial photographic lenses in the spectrograph. Although constrained by the challenging high spectral resolution requirement, we  realize a 40\% reduction in cost per fiber at constant etendue relative to, e.g., DESI. As the reduction of cost is much more significant than the reduction in the amount of light received per fiber, replicating such a system many times is more cost effective than building a single large spectrograph that achieves the same survey speed. We present the design of the prototype telescope and instrument system and the study of its cost effectiveness.
\end{abstract}


\section{Scientific and Technical Motivation}
\label{sec:intro}  

Understanding the rise and fall in the star formation rate within galaxies is essential for a complete picture of galaxy evolution. Recent advances in multi-object integral field spectroscopic instrumentation have made IFS surveys of thousands of nearby galaxies possible, e.g., SAMI\cite{Croom12} and MaNGA\cite{Bundy15}. These data have firmly established that many scaling relations seen for integrated properties of galaxies persist down to $\sim 1$~kpc scales, such as the relations between stellar mass surface density, star formation rate surface density, and gas-phase metallicity\cite{sanchez20}. These data have also established that some regions of galaxies stop their star formation in a relatively rapid manner (referred to as `quenching') to form an underpopulated ``green valley.'' To fully understand what gives rise to these scaling relations as well as to identify quenching mechanisms, we need to connect these `macro' phenomena on kpc scales with  the physical processes of star formation and baryon cycling that occur on even smaller scales. The relevant physical scales start at the size of molecular clouds (\textit{tens} of parsecs), and proceed to the size of star-forming regions at \textit{tenths} of pc. Comprehensive optical spectroscopic observations covering from 0.1~pc to kpc scales are challenging with current instruments. 

There is overwhelming observational and theoretical evidence that stellar feedback is critical for regulating star formation, for driving galactic winds, and for determining the properties of the ISM. However, the detailed physical processes involved with this feedback are far from clear. For example, although we know turbulence plays a major role in keeping star formation inefficient, we do not know how turbulence is driven and on what scale(s) turbulent energy is injected. To tackle this problem requires measuring the power spectra of velocity and emissivity in thin velocity slices.  This in turn requires instruments capable of delivering medium spectral resolution ($R>15,000$) while covering spatial scales from sub-pc to 100 pc \cite{Medina14}. We also have not confirmed observationally the relative importance of photoionization and radiation pressure in driving outflows from star-forming regions. To do so requires 20~km/s separation between different velocity components \cite{Raskutti17, KimKO18}, again requiring medium spectral resolution optical spectroscopy over the same spatial scales. Covering the required spatial and spectral scales is very difficult with current instruments.

Most integral field spectrographs on large telescopes today do not have large fields of view. For example, VLT/MUSE has a field of view of 1 arcmin$^2$. HII regions in the Milky Way are too large on the sky for them to cover, while those in other galaxies are usually too small to resolve at the 0.1-100 pc scale in gound-based seeing. Local group galaxies, like M31, M33, LMC, and SMC, are at about the right distance, however, they still spread over degrees on the sky, making it extremely time-consuming to cover them completely. Moreover, most integral field spectrographs cannot reach spectral resolutions of $R\sim15,000$ (KCWI\cite{Morrissey18}, MEGARA\cite{Garcia-Vargas20} on the GTC, WEAVE\cite{Dalton18} on the WHT, and WIYN IFUs\cite{Bershady05,Eigenbrot18} are some exceptions). The instrument we describe here fills a niche to enable observations over large areas at medium spectral resolution.

We consider the science case here for nebular spectroscopy of Milky Way HII regions. In this regime 0.1~pc physical resolution corresponds to tens of arcseconds on the sky. Consequently we do not need the large plate scales afforded
by large telescopes. Compared to the telescope aperture, more important is the focal ratio of our observing system. Like the aperture setting for photography, fast focal ratios concentrate more light into each pixel, allowing fainter features to be detected in less time. It is extremely challenging to reduce the focal ratio of optics as it leads to degradation of image quality and mismatch with the numerical aperture of optical fibers. However, one can significantly reduce the effective focal ratio of the system by making many identical copies of a telescope and then stacking the data together, as done by the Dragonfly Telephoto Array \cite{AbrahamvD14}. Applying this strategy to spectroscopy requires the replication of many spectrographs. Most astronomical spectrographs cost millions of dollars to build, and are prohibitively expensive to replicate in large numbers. The VIRUS spectrograph \cite{Hill14} has demonstrated modular designs and massive replication can significantly reduce cost and improve cost-effectiveness. Our instrument design takes cost reduction even further via several strategies, while pushing for higher spectral resolution than typical for survey spectrographs. 

The key to achieving this higher-resolution capability at very low cost is to take advantage of a unique combination of thinner fibers, smaller-pixel CMOS detectors, and commercial photographic lenses. Thinner optical fibers ($50~\mu$m core) require shorter collimator focal length for fixed spectral resolution. A shorter focal length reduces the beam width of the spectrograph, allowing the use of off-the-shelf photographic lenses in the spectrograph, cutting cost significantly. It also reduces the size of the grating, cutting cost further. At the same time, thinner fibers enable the use of small pixel commercial CMOS sensors without requiring a large de-magnification factor. 

There have been significant technology advances in both CMOS sensors and photographic lenses, making them more suitable than ever for adoption in astronomical instruments. Today's back-illuminated CMOS sensors have comparable sensitivity to top-level CCDs ($> 90\%$ peak efficiency) and have even lower read noise ($\sim1$ $e^-$), but are much cheaper due to industry-scale mass production. The advance in CMOS also indirectly made commercial lenses more suitable for adoption in spectrographs. The fast readout speed of CMOS has eliminated the need for a flip mirror in commercial cameras. This reduced the back-focus requirement for commercial lenses, enabling them to be designed with faster focal ratio (e.g., f/0.95) and have better image quality. In addition, advanced nano-crystal coating technology gives these commercial lenses high throughput and low scattered-light, enabling the use of more elements to correct for aberrations. 





In this proceeding, we describe the conceptual design of an innovative, high spectral resolution, cost-effective integral field spectrograph. When paired with small telescopes like a commercial Canon telephoto lens, it is ideal for observing star-forming regions in the Milky Way and Local Group galaxies, and for conducting a wide area contiguous IFS mapping of the sky.

\section{Telescope and Spectrograph Design}

\subsection{Design Optimization to Lower Cost} \label{sec:design_optimization}

The entire instrument includes a Canon Telephoto lens serving as a telescope, a double-arm spectrograph, and a fiber-feed that connects them. Because the telescope is small and much cheaper than the spectrograph, our design considerations are very different from those guiding traditional astronomical instruments. When telescopes are large and expensive, it is most cost-effective to make the instrument accept all the light collected by the telescope; hence the telescope sets the system aperture stop. In contrast, since we do not need large telescopes for nebular spectroscopy, the spectrograph is {\it relatively} more expensive. In this case, cost-effective designs sets the system aperture by the spectrograph.

\begin{figure}
    \centering
    \includegraphics[width=0.6\textwidth]{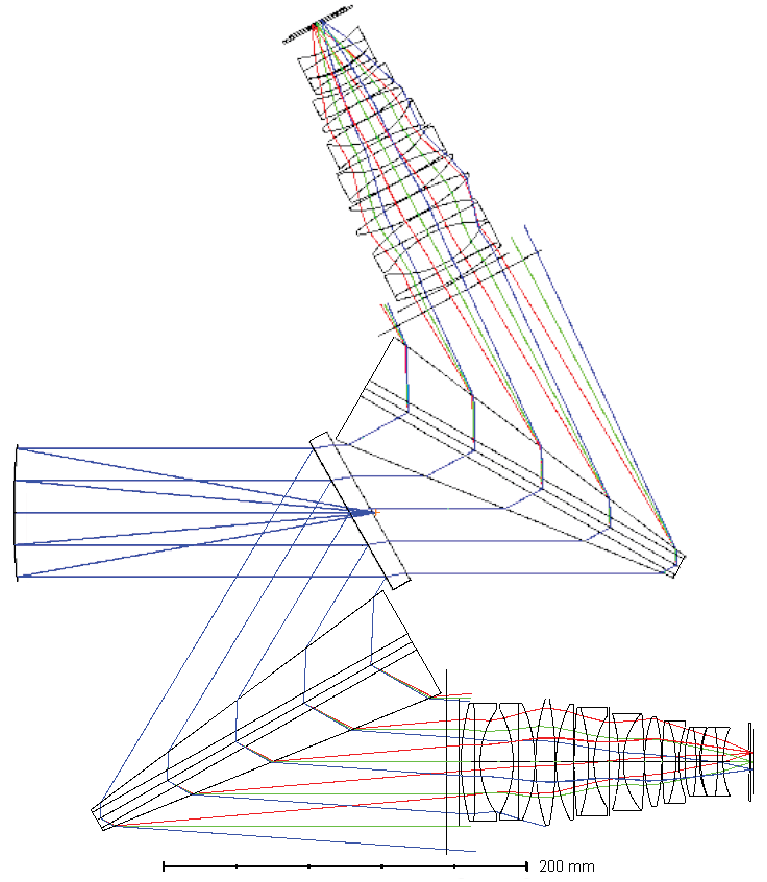}
    \caption{Optical model of the AMASE-P spectrograph. }
    \label{fig:optical_model}
\end{figure}
    
\begin{figure}
\centering
    \includegraphics[width=0.6\textwidth]{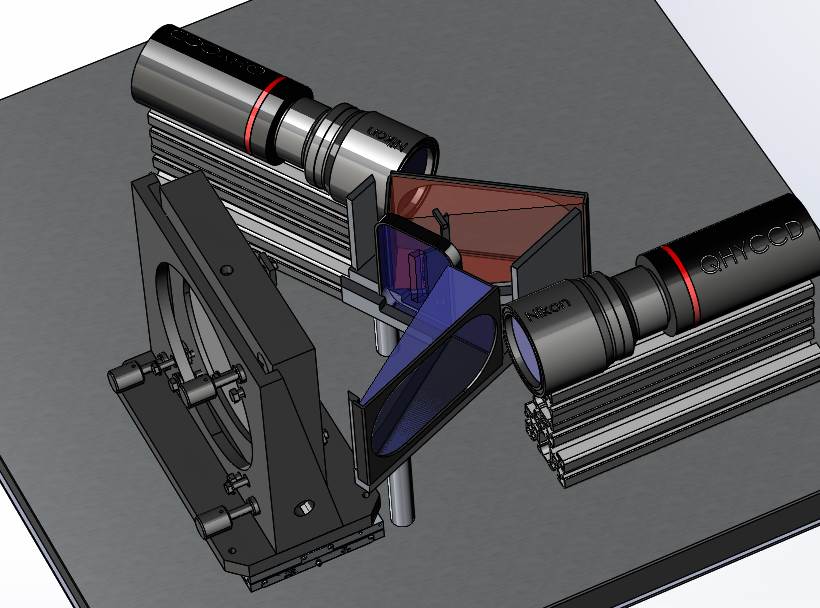}
    \caption{Optomechanical layout of the system. The plates around the grisms are for the blocking of stray light.}
    \label{fig:optomechanical}
\end{figure}

The design of the spectrograph is driven by the need for moderate spectral resolution ($R = \lambda/\Delta\lambda = 15,000$) and large étendue (area $\times$ solid-angle $\times$ efficiency, or $A\Omega\epsilon$) at low cost. Our approach is to adopt small spectrograph entrance apertures to enable the use of commercial optics and CMOS detectors. While we sacrifice system étendue per unit in doing so, {\it significant cost-reduction allows us to more than recoup via spectrograph multiplex.}

The design optimization starts with setting the beam width of the spectrograph by finding the most optimal commercial optics to use as the spectrograph camera.
The most optimal commercial lens for our application is one that has the highest étendue and the highest image quality. The recently released Nikkor 58mm f/0.95 S Noct lens by Nikon in 2019 is the champion in both aspects. It has an amazingly fast focal ratio of f/0.95 and is designed for full-frame detectors ($36{\rm mm} \times 24 {\rm mm}$), meaning it has the largest étendue among all commercial photographic lenses. It is also specifically designed to minimize aberrations for bright point-sources in dark backgrounds, perfect for astrophotography. ARNEO nano-crystal coatings yield high throughput and minimize ghosts and flares. We have measured its throughput and verified the image quality through lab tests and ray-tracing calculations. 
As shown in Fig.~\ref{fig:optical_model}, the lens has 17 elements in 10 groups, including 4 extreme low dispersion elements and 3 aspherical surfaces. This level of sophistication would be extremely challenging and expensive to achieve with custom optics.

This Nikon lens has an entrance pupil of 61~mm, setting the spectrograph beam width. To match the small-pixel CMOS detectors, we adopt the thinnest available multi-mode broad-spectrum optical fiber (Polymicro) with a 50~${\rm \mu m}$-diameter core. With these parameters fixed, the setting of the collimator focal length requires a compromise. On one hand, to maximize throughput and minimize focal ratio degradation requires the fibers are fed with a fast focal ratio close to their numerical aperture. At fixed beam width, this means we want to minimize the focal length of the collimator. On the other hand, for fixed fiber core diameter, the spectral resolution requirement drives a longer collimator focal length to ease the angular dispersion requirement of the grating. We compromise between these two factors and choose a collimator with 200mm focal length. This compromise means the collimated beam is larger than the entrance pupil of the Nikon lens, leading to a significant amount of vignetting. The resulting demagnification factor is 3.45 -- the image of a fiber would have a FWHM of 14.5~${\rm \mu m}$ on the detector, corresponding to 3.86 pixels for the detector we choose.

To sample HII regions in the Milky Way to sub-parsec scales, we choose the $f=400$mm f/2.8 Canon telephoto lens as the telescope. While this overfills the spectrograph pupil, this is not a concern because {\it the cost of the system is not limited by the telescope but by the spectrograph}. The camera optics (Nikkor lenses) are the limiting factor in the system étendue; because our beam size is below a break in the grating cost curve, we place the spectrograph pupil near the front of the camera to maximize system efficiency, rather than on the grating. 


We describe core spectrograph components, telescope, and fiber-feed system below. Figure~\ref{fig:optical_model} shows the optical and  optomechanical models with design parameters summarized in Table 1. 

\begin{table}[]
    \centering
    \caption{Nominal Instrument Design Parameters}
    \begin{tabular}{l|c|l}
        \hline
        Item & Value & Notes\\\hline
        Telescope & D=143mm f/2.8 & Canon 400~mm f/2.8 L~III IS USM \\
        {\bf Spatial resolution} & 25.8 arcsec & fiber core:clad:buffer 50:70:85~${\rm \mu m}$  \\
        Fiber IFU fill factor & 31.4\% & 85~${\rm \mu m}$ center-to-center spacing\\
        Number of Fibers  & 547 & per spectrograph \\
        Simultaneous field coverage & 79 arcmin$^2$ & per spectrograph \\
        On-sky footprint & 16 arcmin $\times$ 16 arcmin & per spectrograph \\
        Fiber-to-fiber spacing on slit & 100~${\rm \mu m}$ & 60~mm pseudo-slit length\\
        Collimator focal length & 200~mm & Silver-coated spherical mirror \\ 
        Camera focal length & 58~mm & Nikkor 58mm f/0.95 S Noct \\
        Detector array & $9576\times6388$ pix & $36~{\rm mm} \times24~$mm\\
        PSF FWHM on detector & 3.86 pixel & pixel size: 3.76~$\mu$m \\
        Red grating density & 2343 l/mm & blazed at 6604~\AA \\
        Red spectral coverage  & 6250-6850~\AA\ ($\epsilon>10\%$ ~of~peak) & 6500-6770~\AA\ ($\epsilon>70\%$ of peak) \\
        Blue grating density & 3156 l/mm & blazed at 4903~\AA \\
        Blue spectral coverage  & 4640-5092~\AA\ ($\epsilon>10\%$ of peak) & 4820-5020~\AA\ ($\epsilon>70\%$ of peak) \\
        {\bf Spectral Resolution ($R$)} & 15,000 & red: @6604~\AA, blue: @4903\AA\ \\
    \hline
    \end{tabular}
    \label{tab:nominalparam}
\end{table}

\subsection{Core Spectrograph Components}\label{sec:components}



{\bf Collimator--Entrance-slit Assembly:} The collimator is a silver-coated spherical mirror with a focal length of 200mm. We align 547 fibers along a curved slit with a radius of curvature of 200mm. The fibers are split into a number of fiber blocks to allow angling fibers to follow the curvature of the slit. Fibers are spaced 100${\rm \mu m}$ center-to-center to minimize cross-talk. Three-fiber spacing gaps between adjacent fiber blocks allow for scattered light measurements. The fibers are fed at f/2.8. The output beam is slightly faster, estimated to be about f/2.7 due to focal ratio degradation. The collimator will form a collimated beam of 74mm diameter.




 
{\bf Dichroic:} Grism tilt angles of 61 degrees are accommodated via a dichroic beam splitter designed for 29 deg incidence angle, reflecting below 576~nm. This optimally positions the blue camera and detector. In order to place the system pupil closer to the blue camera to reduce vignetting, we place the dichroic as close to the collimator as possible. This configuration requires a slot in the  dichroic, which is located a few mm in front of the slit, to allow light from the slit to pass through. The slot width  is minimized to be just larger than the thickness of the slit plate. A representative quote yields reflectance for both polarizations in the blue above 99\%, and  transmittance in the red above 98\%. 

{\bf VPH grism:} The VPH-grism is a critical design element for achieving the high spectral resolution necessary for our science goals. Volume-phase holographic transmission gratings deliver high diffraction efficiency and optimum geometry. To achieve the desired spectral resolution we adopt a $\sim35.3\deg$ incidence angle in the grating medium of both channels, optimized to reach peak diffraction efficiency for both polarizations simultaneously. Line-densities and super-blaze peaks are given in Table 1. We bracket the gratings with two small-angle, fused-silica prisms to further increase the angular dispersion to reach the science requirement. The expected diffraction efficiency is $\sim$90\% at the peak and $>$ 70\% over a $>200$\AA\ window in the blue and $>270$\AA\ window in the red. 

{\bf Camera Optics:} The adopted Nikkor lens has excellent throughput, which we have measured by comparing it to uncoated fused silica lens. Its absolute throughput is $>$90\% between 550nm and 695nm and $>$ 80\% between 462nm and 732nm. It does have significant vignetting for large field angles. The collimated beam has a 74mm diameter, larger than the 61mm entrance pupil. This overfilling helps reduce the vignetting of beams at large field angles relative to the on-axis beam. One can argue that our on-axis beam has significant vignetting given the f/2.8 input beam at the telescope. Instead of choosing a shorter focal length for the collimator, we choose to sacrifice throughput in order to reach the high spectral resolution required by the science. 


\begin{figure}
    \centering
    \includegraphics[width=\textwidth]{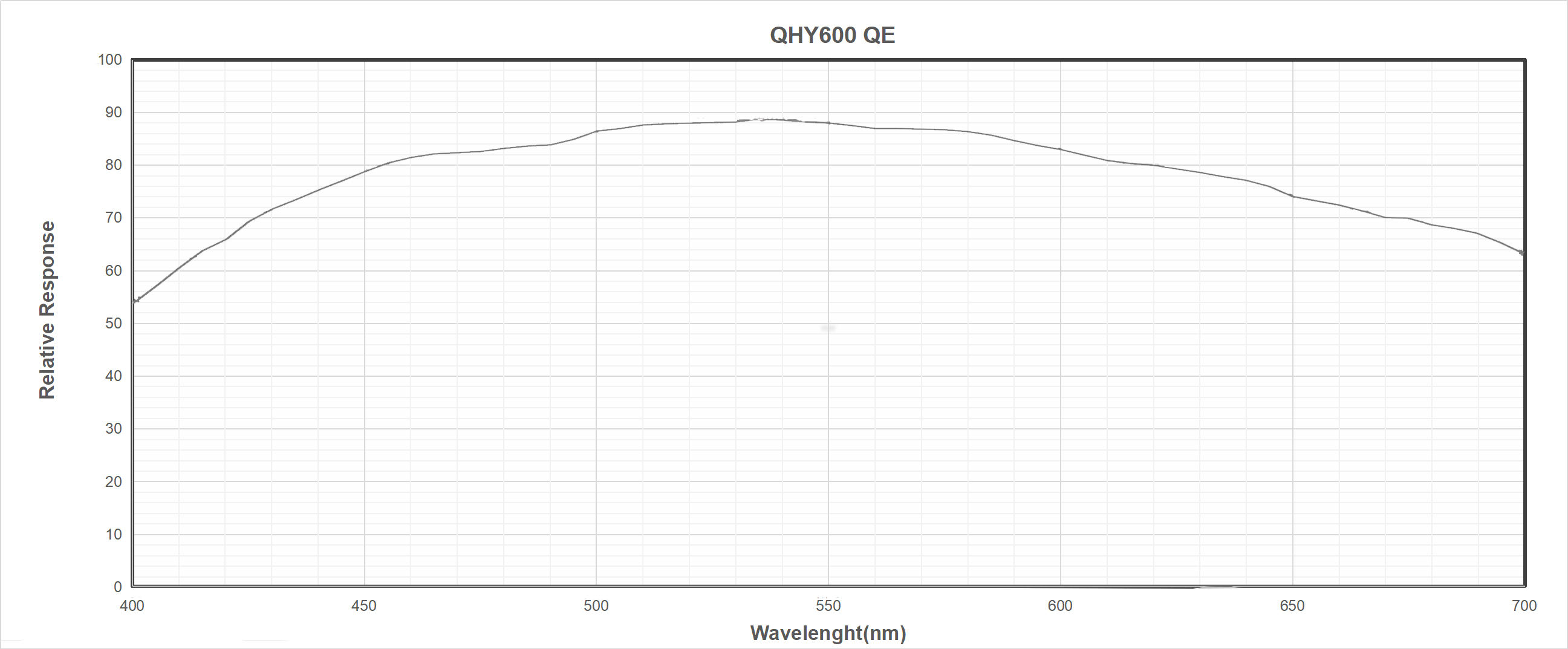}
    \caption{Absolute quantum efficiency of the IMX455 measured by QHYCCD. Figure adapted from 
    https://www.qhyccd.com/index.php?m=content\&c=index\&a=show\&catid=94\&id=55\&cut=1 . Also see this page for a discussion: https://www.qhyccd.com/index.php?m=content\&c=index\&a=show\&catid=23\&id=261}
    \label{fig:cmosqe}
\end{figure}

{\bf Detector Assembly}: The detector package is QHY600M monochromatic camera from QHYCCD. It uses Sony's state-of-the-art, back-illuminated, 16-bit full-frame CMOS sensor, IMX455. Quantum efficiency is about 90\% at the peak and above 70\% for the whole wavelength range we use (Figure~\ref{fig:cmosqe}). Read-noise is 1.68 e- for a full well depth of 21.7k e-, or 1.39e- for a full well depth of 3k e- \footnote{https://www.qhyccd.com/index.php?m=content\&c=index\&a=show\&catid=94\&id=55\&cut=1}. The dark current is 0.0022 e-/pix/s at T=-20C,  dropping by a factor of two for every $\Delta$T = -10C. 
A water-cooled version at the same price is available that can bring the detector to -45 degree below ambient temperature. The front window of the camera is heated to prevent condensation. We will replace the front window of the detector by a custom plano-concave lens to flatten the field curvature. No physical shutter is needed for the camera as it has an electronic rolling shutter. 

\begin{figure}
\vspace{-20pt}
    \centering
    \includegraphics[width=0.6\textwidth]{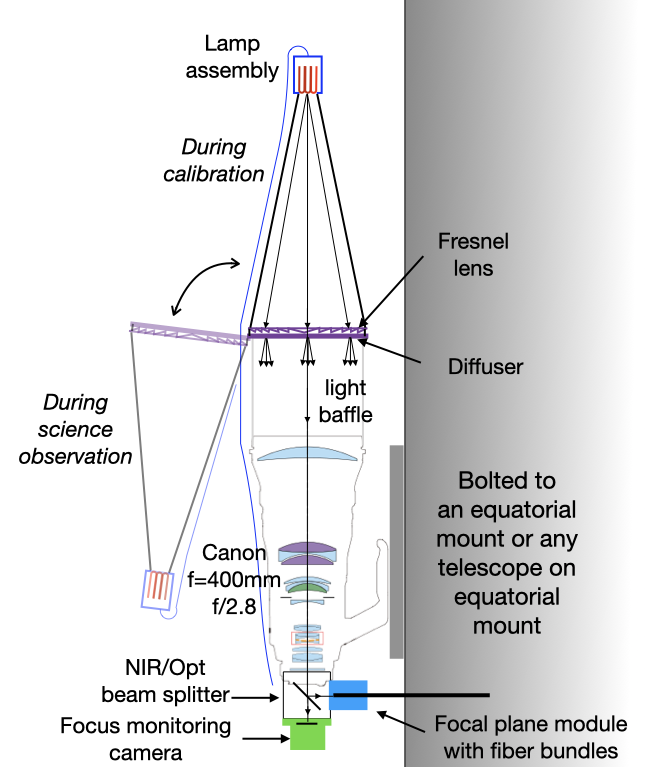}
    \caption{Calibration system and focusing mechanism conceptual design. The Canon telephoto lens can be independently mounted or bolted to a small telescope on an equatorial mount (as shown here). The cone on top of the telephoto lens is the calibration system. During science observations, the cone swings off. {\it This allows us to calibrate at any telescope position to correct for flexure induced throughput and line spread function variation.}}
    \label{fig:telescope_schematic}
\end{figure}






\section{Telescope, Focal Plane and Fiber-Feed System}

The telescope is a $f=400$~mm f/2.8 Canon telephoto lens feeding a hexagonal fiber bundle of 547 fibers. The telescope will be mounted on an equatorial mount to avoid field rotation. The 50${\rm \mu m}$ fiber core diameter corresponds to 26 arcsec on sky. Excluding gaps between the fibers, the bundle covers an area of 80 arcmin$^2$.  The bundle will be packaged in a focal-plane assembly that includes focus and acquisition camera described below. The IFU fibers route in a cable to the slit assembly (\S\ref{sec:components}) in the spectrograph. The fiber cable is designed and jacketed for the required routing and actuation. 
We have identified a suitable vendor who can fabricate the ferrule and fiber blocks for our smaller fibers using a photolithographic process necessary for these micro-precision components. Specific design considerations include the following elements.




{\bf IFU feed}:
Fiber telecentricity is required to maximize throughput. 
We have measured the position of the exit pupil for the Canon telephoto lens in the lab and found it is $183.5\pm1 {\rm mm}$ away from the focal plane.  We plan to cement an AR coated plano-convex lens on the fiber bundle surface to make the telecentric correction and to minimize Fresnel losses. 


{\bf Focus and guiding}: We defocus the telescope image slightly to smooth radial light distributions within fibers;  this removes the dependence of the line spread function on the intrinsic light distribution within a fiber footprint and reduces the impact of  differential atmospheric refraction for stars on fiber edges. Using a rented lens, we found the slightly-defocused PSF for the Canon 400mm f/2.8 lens is smooth with no sharp boundaries. 
To keep constant defocus, we implement a monitoring mechanism via a 45 deg optical/near-IR dichroic beamsplitter placed before the focal plane (Figure~\ref{fig:telescope_schematic}).  Wavelengths bluer than 750~nm are reflected to the fiber bundle; redder light is transmitted to a focus-monitoring CMOS sensor. We adjust the relative positioning of the fiber bundles and the CMOS sensor so that when a star is in focus in the focus-monitoring sensor, it has a 26 arcsec FWHM on the fiber bundle. This adjustment will be done in the lab using a microscope looking towards the beamsplitter at the bundle and the sensor, which will also provide the relative registration for astrometry. Then we will test it using an illuminated pinhole placed several meters away. By scanning the pinhole across the bundle and the sensor, we can measure the profile of the pin hole image on the fiber bundle. The relative positions will be locked and checked for repeatability. 
During commissioning, by scanning across a bright star, we can verify the PSF on the focal plane. During observations, we will only need to keep the star focused in the focus monitoring camera and we will be sure we have the desired focus offset for the IFU.  The transmitted images through the dichroic beam splitter will have astigmatism. This is useful for distinguishing the front and back side of focus. The electronic focus control will be provided by Birger Engineering.



\section{Calibration System}
 Our science goals require accurate measurement of line width (to measure gas turbulence and kinematic temperature) and line ratios (to measure extinction and abundances). This requires  flat/arc calibration at any pointing position to take out the flexure-induced throughput and line spread function variations. Figure~\ref{fig:telescope_schematic} illustrates the design of the calibration system. It consists of a cone mounted on top of the telephoto lens. At the top of the cone is the lamp assembly including a quartz lamp, two arc lamps (Ne, Th/Ar), and a diaphragm controlling the light output. At the base of the cone, we place a Fresnel lens and a ground glass diffuser. With the lamps placed at the focus of the Fresnel lens, the diffuser is illuminated with a collimated beam. With a coarse grit, the ground glass diffuser provides a wide diffusing beam with flat intensity over the central few degrees, sufficient to completely cover the IFU footprint. When not in use, the cone swings out of the beam and rests on the side of the lens. The Fresnel lens will be slower than f/2 to ensure sufficiently uniform pupil illumination. 



\section{Preliminary Tests of the Design}

\subsection{Throughput of the Nikkor lens}

\begin{figure}
    \centering
    \includegraphics[width=0.6\textwidth]{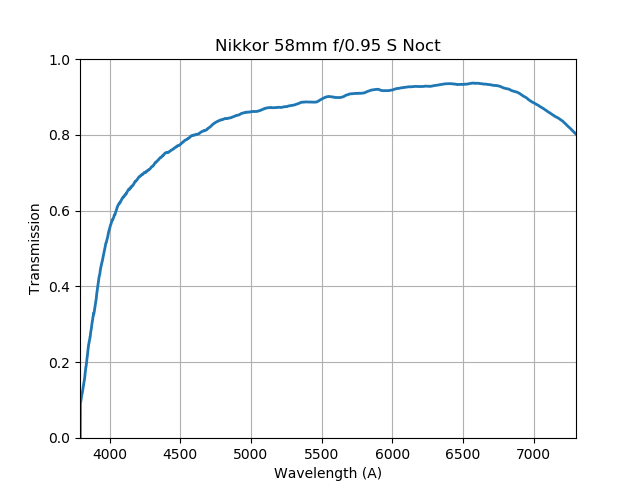}
    \caption{Absolute throughput measurement of the Nikkor 58mm f/0.95 S Noct lens.}
    \label{fig:throughput}
\end{figure}

We have measured the absolute throughput of the Nikkor lens by comparing it to a reference optical system constructed using uncoated fused silica lenses. The reference system is made to have the same focal length as the Nikkor lens. Both systems are used to image the spectrum of a quartz halogen lamp placed a few meters away, through a 600 l/mm transmission grating from Thorlabs. An adjustable iris is placed in front of the lenses to minimize intrinsic vignetting of the systems. A Pen-Ray HgNe lamp is used to provide wavelength calibration. The reference system is composed of two uncoated fused silica plano-convex lens whose 4 air-glass surfaces will yield an absolute throughput of 86.7\%. The measured absolute throughput of the Nikkor lens is shown in Figure~\ref{fig:throughput}.

\subsection{Image Quality}
We have simulated the expected image quality using Zemax. The resulting line widths satisfy our resolution requirements. The spherical aberration introduced by the spherical collimator can be adequately compensated for by a very slight adjustment of the Nikkor lens focus. We have verified this in the lab  using a 50$\mu$m core fiber, a spherical collimator, and a rented Nikon lens. The result is totally consistent with our simulations. With the addition of the grism, the images at large slit height are diagonally elongated. This can be accommodated in the analysis pipelines for spectral extraction and estimating instrumental resolution, using spectral-perfectionism algorithm \cite{BoltonS10}.

\subsection{Thin-fiber positioning}
\begin{figure}
    \begin{center}
    \includegraphics[width=0.6\textwidth]{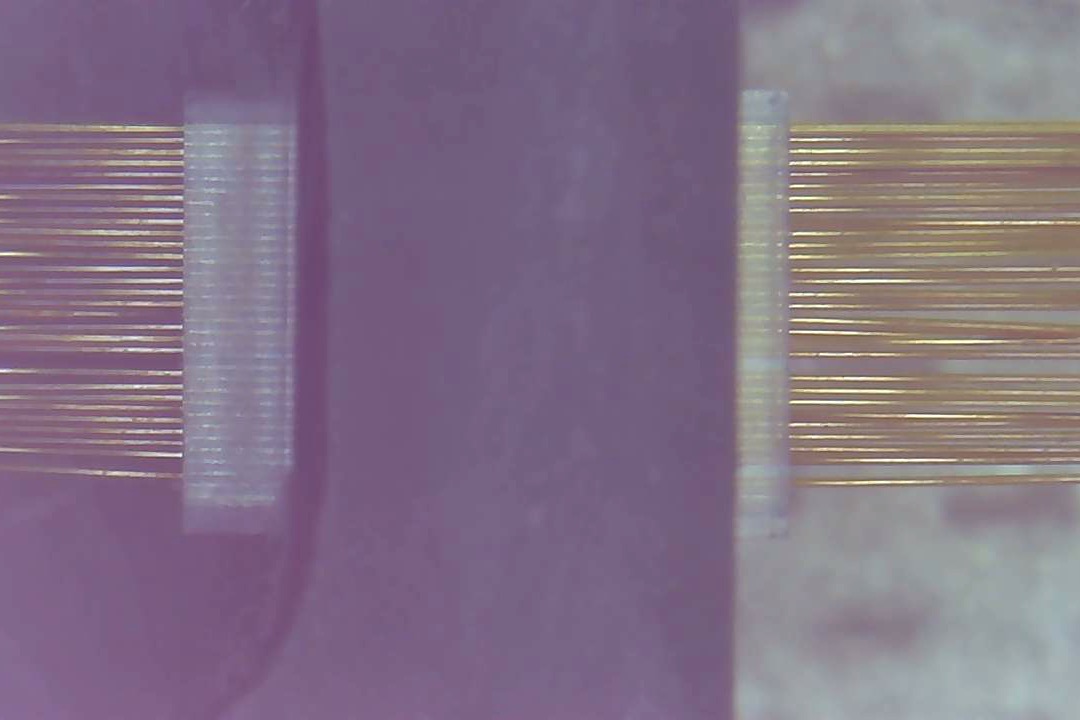}
    \end{center}
    \caption{Fused silica fiber blocks populated with fibers of 85${\rm \mu m}$ OD and 50${\rm \mu m}$ core. The block is clamped by a mounting fixture so the central part is obscured. }
    \label{fig:fiberblocks}
\end{figure}

The 50$\mu$m core fibers we are using are more challenging to handle and align accurately on a pseudo-slit or IFU using the technology developed for the larger MaNGA fiber system. For example, wire-EDM technology is impractical for cutting V-grooves with radius of curvature smaller than 40$\mu$m required to precisely seat fibers with 85$\mu$m OD. Thus, we worked with a commercial fabricator (FEMTOprint) to develop a process combinning femto-pulse laser drilling and photolithographic etching. This effort resulted in fused-silica blocks with cylindrical apertures arrayed in a pseudo-slit. Each aperture has 1$\mu$m centration and diameter tolerance. We have acquired test blocks, and populated them with $50{\rm \mu m}$-core fiber. The test was largely successful: Fibers were smoothly inserted in most cases. However, the sharp edges at the aperture entry locations strip buffer, causing fibers to become excessively brittle. We have modified the block design to include chamfers at the aperture edges, and a staggered pseudo-slit to accommodate the chamfers while delivering the same efficient packing density as in our initial design. We expect the next version to work without problem. Figure~\ref{fig:fiberblocks} shows a picture under microscope for such a fiber block with $33 \times 50{\rm \mu m}$ fibers inserted. 

\section{Expected Sensitivity}\label{sec:sensitivity}

\begin{table}[]
    \centering
    \begin{tabular}{c|c|c|c|c|c|c|c|c}
    \hline\hline
        Channel & Collimator & Dichroic & Prisms & Grating & Camera & Detector & Vignetting & Overall \\\hline
        Blue & 93\% & 99\% & 94\% & 91\% & 86\% & 84\% & 50\% & 28.2\% \\
        Red  & 97\% & 98\% & 94\% & 91\% & 93\% & 72\% & 51\% & 27.4\%\\
        \hline
    \end{tabular}
    
    \caption{Expected system throughput for the central wavelength in each channel. Vignetting is separately accounted for and discussed in \S\ref{sec:sensitivity}.}
    \label{tab:efficiency}
\end{table}

As listed in Table~\ref{tab:efficiency}, the expected throughput of our instrument is about 28\% and 27\% for the central wavelengths in the blue and red channels, respectively. The vignetting is a significant hit to the overall throughput. Part of it comes from the slot in the dichroic beamsplitter and the obstruction by the slit plate. But most of the vignetting is due to a design choice we made in order to reach the high spectral resolution (see \S\ref{sec:design_optimization}). If we reduce the focal length of the collimator, we could significantly improve the throughput, but we will not reach the required spectral resolution. 

\begin{figure}
    \centering
    \includegraphics[width=0.45\textwidth]{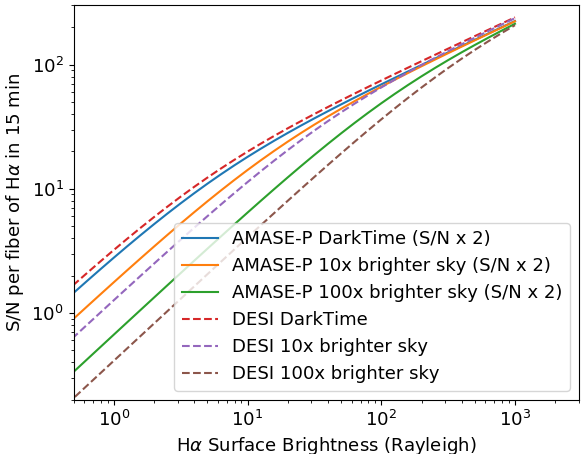}
    \caption{Expected S/N in \hal\ per fiber for AMASE-P and DESI with a single 15-min exposure, in 3 different background conditions. The S/N for AMASE-P is doubled for ease of comparison with DESI.}
    \label{fig:surveyspeed}
\end{figure}

It is convenient to compare the etendue of the AMASE-P spectrograph with the DESI spectrograph \cite{Perruchot18}, which is a highly-optimized, state-of-the-art broad band-pass system. DESI has an f/3.86 input, and the throughput at these wavelengths is measured to be about 64\% and 66\% \cite{Perruchot18}. Given our smaller fiber core diameter (50 ${\rm \mu m}$) compared to DESI's 107 ${\rm \mu m}$ fiber cores, our overall etendue per fiber (fiber input beam solid angle $\times$ fiber area $\times$ spectrograph throughput) is 18\% of that for DESI. However, our hardware costs also differ dramatically. Each DESI spectrograph has 500 fibers and cost \$1.5M  \cite{MegaMapper_whitepaper}, while our prototype instrument, AMASE-P, will have 547 fibers and cost \$184k for the spectrograph hardware. Our per fiber cost is 11\% of that for DESI.  {\it AMASE-P is about 1.6 times more cost effective per unit etendue at 4 to 5 times higher spectral resolution but 18\% of the passband coverage.} 
To be clear, AMASE-P is not a panacea; DESI delivers 5.4 times higher bandwidth and 23\% more spectral resolution elements. However, for the specific emission-line survey targeted by AMASE-P, our design offers significant advances in {\it lowering} cost while {\it increasing} spectral resolution.

AMASE-P's telescope and fiber system are expected to yield a throughput of 76\% and 81\% in the blue and red channels, respectively. Figure~\ref{fig:surveyspeed} shows the expected S/N as a function of \hal\ surface brightness (assuming an unresolved line width), under different conditions for DESI and AMASE-P. This also takes into account the sky background, detector read noise and dark current. The AMASE-P curves are scaled up by a factor of 2 for ease of comparison. The higher spectral resolution allows AMASE-P to be relatively faster in brighter time, further increasing cost-effectiveness for the narrow emission-line science. We obtain a S/N of 3 in $5\times15$min exposures for an \hal\ surface brightness of 1 Rayleigh.

\section{Summary}
Here we have presented the conceptual design of a cost-effective, high spectral resolution, integral field spectrograph capable of reaching spectral resolution of up to R = 15,000. In principle, it can be used on both small and large telescopes to enable wide-area, integral field, spectroscopic surveys. On small telescopes with high-angle VPH grisms, it is ideal for covering and \textit{resolving} key strong-line diagnostics across large HII regions in the Milky Way and mapping all components of the interstellar medium in nearby galaxies. The cost-effectiveness of the design is amenable to lower-angle VPH grism applications that would yield full spectral coverage between 400-800 nm for study of stellar populations in integrated light as well as intensity mapping.

\acknowledgments 


\begin{thebibliography}{10}

\bibitem{Croom12}
{Croom}, S.~M., {Lawrence}, J.~S., {Bland-Hawthorn}, J., {Bryant}, J.~J.,
  {Fogarty}, L., {Richards}, S., {Goodwin}, M., {Farrell}, T., {Miziarski}, S.,
  {Heald}, R., {Jones}, D.~H., {Lee}, S., {Colless}, M., {Brough}, S.,
  {Hopkins}, A.~M., {Bauer}, A.~E., {Birchall}, M.~N., {Ellis}, S., {Horton},
  A., {Leon-Saval}, S., {Lewis}, G., {L{\'o}pez-S{\'a}nchez}, {\'A}.~R., {Min},
  S.-S., {Trinh}, C., and {Trowland}, H., ``{The Sydney-AAO Multi-object
  Integral field spectrograph},'' {\em MNRAS}, {\bf 421},  872--893 (Mar.
  2012).

\bibitem{Bundy15}
{Bundy}, K., {Bershady}, M.~A., {Law}, D.~R., {Yan}, R., {Drory}, N.,
  {MacDonald}, N., {Wake}, D.~A., {Cherinka}, B., {S{\'a}nchez-Gallego}, J.~R.,
  {Weijmans}, A.-M., {Thomas}, D., {Tremonti}, C., {Masters}, K., {Coccato},
  L., {Diamond-Stanic}, A.~M., {Arag{\'o}n-Salamanca}, A., {Avila-Reese}, V.,
  {Badenes}, C., {Falc{\'o}n-Barroso}, J., {Belfiore}, F., {Bizyaev}, D.,
  {Blanc}, G.~A., {Bland-Hawthorn}, J., {Blanton}, M.~R., {Brownstein}, J.~R.,
  {Byler}, N., {Cappellari}, M., {Conroy}, C., {Dutton}, A.~A., {Emsellem}, E.,
  {Etherington}, J., {Frinchaboy}, P.~M., {Fu}, H., {Gunn}, J.~E., {Harding},
  P., {Johnston}, E.~J., {Kauffmann}, G., {Kinemuchi}, K., {Klaene}, M.~A.,
  {Knapen}, J.~H., {Leauthaud}, A., {Li}, C., {Lin}, L., {Maiolino}, R.,
  {Malanushenko}, V., {Malanushenko}, E., {Mao}, S., {Maraston}, C.,
  {McDermid}, R.~M., {Merrifield}, M.~R., {Nichol}, R.~C., {Oravetz}, D.,
  {Pan}, K., {Parejko}, J.~K., {Sanchez}, S.~F., {Schlegel}, D., {Simmons}, A.,
  {Steele}, O., {Steinmetz}, M., {Thanjavur}, K., {Thompson}, B.~A., {Tinker},
  J.~L., {van den Bosch}, R.~C.~E., {Westfall}, K.~B., {Wilkinson}, D.,
  {Wright}, S., {Xiao}, T., and {Zhang}, K., ``{Overview of the SDSS-IV MaNGA
  Survey: Mapping nearby Galaxies at Apache Point Observatory},'' {\em ApJ}, {\bf 798},  7 (Jan. 2015).

\bibitem{sanchez20}
{S{\'a}nchez}, S.~F., ``{Spatially Resolved Spectroscopic Properties of
  Low-Redshift Star-Forming Galaxies},'' {\em ARAA},{\bf 58},  99--155 (Aug.
  2020).

\bibitem{Medina14}
{Medina}, S. N.~X., {Arthur}, S.~J., {Henney}, W.~J., {Mellema}, G., and
  {Gazol}, A., ``{Turbulence in simulated H II regions},'' {\em MNRAS}, {\bf
  445},  1797--1819 (Dec. 2014).

\bibitem{Raskutti17}
{Raskutti}, S., {Ostriker}, E.~C., and {Skinner}, M.~A., ``{Numerical
  Simulations of Turbulent Molecular Clouds Regulated by Radiation Feedback
  Forces. II. Radiation-Gas Interactions and Outflows},'' {\em ApJ}, {\bf 850},
   112 (Dec 2017).

\bibitem{KimKO18}
{Kim}, J.-G., {Kim}, W.-T., and {Ostriker}, E.~C., ``{Modeling UV Radiation
  Feedback from Massive Stars. II. Dispersal of Star-forming Giant Molecular
  Clouds by Photoionization and Radiation Pressure},'' {\em ApJ}, {\bf 859},
  68 (May 2018).

\bibitem{Morrissey18}
{Morrissey}, P., {Matuszewski}, M., {Martin}, D.~C., {Neill}, J.~D., {Epps},
  H., {Fucik}, J., {Weber}, B., {Darvish}, B., {Adkins}, S., {Allen}, S.,
  {Bartos}, R., {Belicki}, J., {Cabak}, J., {Callahan}, S., {Cowley}, D.,
  {Crabill}, M., {Deich}, W., {Delecroix}, A., {Doppman}, G., {Hilyard}, D.,
  {James}, E., {Kaye}, S., {Kokorowski}, M., {Kwok}, S., {Lanclos}, K.,
  {Milner}, S., {Moore}, A., {O'Sullivan}, D., {Parihar}, P., {Park}, S.,
  {Phillips}, A., {Rizzi}, L., {Rockosi}, C., {Rodriguez}, H., {Salaun}, Y.,
  {Seaman}, K., {Sheikh}, D., {Weiss}, J., and {Zarzaca}, R., ``{The Keck
  Cosmic Web Imager Integral Field Spectrograph},'' {\em ApJ}, {\bf 864},  93
  (Sept. 2018).
  
\bibitem{Garcia-Vargas20}
{Garc{\'\i}a-Vargas}, M.~L., {Carrasco}, E., {Moll{\'a}}, M., {Gil de Paz}, A.,
  {Berlanas}, S.~R., {Cardiel}, N., {G{\'o}mez-Alvarez}, P., {Gallego}, J.,
  {Iglesias-P{\'a}ramo}, J., {Cedazo}, R., {Pascual}, S., {Castillo-Morales},
  A., {P{\'e}rez-Calpena}, A., and {Mart{\'\i}nez-Delgado}, I., ``{MEGARA-GTC
  stellar spectral library: I},'' {\em MNRAS}, {\bf 493},  871--898 (Mar. 
  2020).
  
\bibitem{Dalton18}
{Dalton}, G., {Trager}, S., {Abrams}, D.~C., {Bonifacio}, P., {Aguerri}, J.
  A.~L., {Vallenari}, A., {Middleton}, K., {Benn}, C., {Dee}, K., {Say{\`e}de},
  F., {Lewis}, I., {Pragt}, J., {Pic{\'o}}, S., {Walton}, N., {Rey}, J.,
  {Allende Prieto}, C., {Lhom{\'e}}, {\'E}., {Terrett}, D., {Brock}, M.,
  {Gilbert}, J., {Ridings}, A., {Verheijen}, M., {Tosh}, I., {Steele}, I.,
  {Stuik}, R., {Kroes}, G., {Tromp}, N., {Kragt}, J., {Lesman}, D., {Mottram},
  C., {Bates}, S., {Gribbin}, F., {Burgal}, J.~A., {Herreros}, J.~M.,
  {Delgado}, J.~M., {Martin}, C., {Cano}, D., {Navarro}, R., {Irwin}, M.,
  {Lewis}, J., {Gonzales Solares}, E., {O'Mahony}, N., {Bianco}, A., {Zurita},
  C., {ter Horst}, R., {Molinari}, E., {Lodi}, M., {Guerra}, J., {Baruffolo},
  A., {Carrasco}, E., {Farkas}, S., {Schallig}, E., {Hill}, V., {Smith}, D.,
  {Drew}, J., {Poggianti}, B., {Pieri}, M., {Jin}, S., {Dominquez Palmero}, L.,
  {Fari{\~n}a}, C., {Martin}, A., {Worley}, C., {Murphy}, D., {Hidalgo}, A.,
  {Mignot}, S., {Bishop}, G., {Guest}, S., {Elswijk}, E., {de Haan}, M.,
  {Hanenburg}, H., {Salasnich}, B., {Mayya}, D., {Izazaga-P{\'e}rez}, R., and
  {Peralta de Arriba}, L., ``{Construction progress of WEAVE: the next
  generation wide-field spectroscopy facility for the William Herschel
  Telescope},'' in [{\em Ground-based and Airborne Instrumentation for
  Astronomy VII}{\nolinebreak\hspace{0.1em}]},  {Evans}, C.~J., {Simard}, L.,
  and {Takami}, H., eds., {\em Society of Photo-Optical Instrumentation
  Engineers (SPIE) Conference Series} {\bf 10702},  107021B (July 2018).

\bibitem{Bershady05}
{Bershady}, M.~A., {Andersen}, D.~R., {Verheijen}, M. A.~W., {Westfall}, K.~B.,
  {Crawford}, S.~M., and {Swaters}, R.~A., ``{SparsePak: A Formatted Fiber
  Field Unit for The WIYN Telescope Bench Spectrograph. II. On-Sky
  Performance},'' {\em ApJS}, ~{\bf 156},  311--344 (Feb. 2005).

\bibitem{Eigenbrot18}
{Eigenbrot}, A. and {Bershady}, M.~A., ``{Vertical Population Gradients in NGC
  891. I. {\ensuremath{\nabla}}Pak Instrumentation and Spectral Data},'' {\em
  ApJ}, {\bf 853},  114 (Feb. 2018).

\bibitem{AbrahamvD14}
{Abraham}, R.~G. and {van Dokkum}, P.~G., ``{Ultra-Low Surface Brightness
  Imaging with the Dragonfly Telephoto Array},'' {\em PASP}, {\bf 126},  55
  (Jan 2014).

\bibitem{Hill14}
{Hill}, G.~J., {Tuttle}, S.~E., {Drory}, N., {Lee}, H., {Vattiat}, B.~L.,
  {DePoy}, D.~L., {Marshall}, J.~L., {Kelz}, A., {Haynes}, D., {Fabricius},
  M.~H., {Gebhardt}, K., {Allen}, R.~D., {Anwad}, H., {Bender}, R., {Blanc},
  G., {Chonis}, T., {Cornell}, M.~E., {Dalton}, G., {Good}, J., {Jahn}, T.,
  {Kriel}, H., {Landriau}, M., {MacQueen}, P.~J., {Murphy}, J.~D., {Peterson},
  T.~W., {Prochaska}, T., {Nicklas}, H., {Ramsey}, J., {Roth}, M.~M., {Savage},
  R.~D., and {Snigula}, J.,  [{\em {VIRUS: production and deployment of a
  massively replicated fiber integral field spectrograph for the upgraded
  Hobby-Eberly Telescope}}{\nolinebreak\hspace{0.1em}]}, vol.~9147 of {\em
  Society of Photo-Optical Instrumentation Engineers (SPIE) Conference Series},
   91470Q (2014).

\bibitem{BoltonS10}
{Bolton}, A.~S. and {Schlegel}, D.~J., ``{Spectro-Perfectionism: An Algorithmic
  Framework for Photon Noise-Limited Extraction of Optical Fiber
  Spectroscopy},'' {\em PASP},{\bf 122},  248 (Feb. 2010).

\bibitem{Perruchot18}
{Perruchot}, S., {Guy}, J., {Le Guillou}, L., {Blanc}, P.~E., {Ronayette}, S.,
  {R{\'e}gal}, X., {Castagnoli}, G., {Sepulveda}, E., {Le Van Suu}, A.,
  {Jullo}, E., {Cuby}, J.~G., {Karkar}, S., {Ghislain}, P., {Repain}, P.,
  {Carton}, P.~H., {Magneville}, C., {Ealet}, A., {Escoffier}, S., {Secroun},
  A., {Cousinou}, M.~C., {Honscheid}, K., {Elliot}, A., {Jelinsky}, P.,
  {Brooks}, D., and {Tarl{\`e}}, G., ``{Integration and testing of the DESI
  multi-object spectrograph: performance tests and results for the first unit
  out of ten},'' in [{\em Ground-based and Airborne Instrumentation for
  Astronomy VII}{\nolinebreak\hspace{0.1em}]},  {Evans}, C.~J., {Simard}, L.,
  and {Takami}, H., eds., {\em Society of Photo-Optical Instrumentation
  Engineers (SPIE) Conference Series} {\bf 10702},  107027K (July 2018).

\bibitem{MegaMapper_whitepaper}
{Schlegel}, D., {Kollmeier}, J.~A., and {Ferraro}, S., ``{The MegaMapper: a
  z\&gt;2 spectroscopic instrument for the study of Inflation and Dark
  Energy},'' in {\em Bulletin of AAS},   {\bf 51},  229
  (Sep 2019).

\end{thebibliography}

\end{document}